\documentclass{emulateapj}
\usepackage{txfonts}

\shorttitle{Properties of M31 Globular Clusters}

\shortauthors{Ma et al.}

\begin{document}
\slugcomment{AJ, in press}
\title{Spectral Energy Distributions and Age Estimates of 39
Globular Clusters in M31}

\author{Jun Ma,\altaffilmark{1}
Zhou Fan,\altaffilmark{1,2} Richard de Grijs,\altaffilmark{3,1}
Zhenyu Wu,\altaffilmark{1} Xu
Zhou,\altaffilmark{1} Jianghua Wu,\altaffilmark{1} Zhaoji
Jiang,\altaffilmark{1} and Jiansheng Chen\altaffilmark{1}}

\altaffiltext{1}{National Astronomical Observatories, Chinese
Academy of Sciences, Beijing, 100012, P. R. China;\\
majun@vega.bac.pku.edu.cn}

\altaffiltext{2}{Graduate University, Chinese Academy of Sciences,
Beijing, 100039, P. R. China}

\altaffiltext{3}{Department of Physics \& Astronomy, The University of
Sheffield, Hicks Building, Hounsfield Road, Sheffield S3 7RH, UK}

\begin{abstract}
This paper supplements \citet{jiang03}, who studied 172 M31 globular
clusters (GCs) and globular cluster candidates from \citet{batt87} on
the basis of integrated photometric measurements in the
Beijing-Arizona-Taiwan-Connecticut (BATC) photometric system. Here, we
present multicolor photometric CCD data (in the BATC system) for the
remaining 39 M31 GCs and candidates. In addition, the ages of 35 GCs
are constrained by comparing our accurate photometry with updated
theoretical stellar synthesis models. We use photometric measurements
from {\sl GALEX} in the far- and near-ultraviolet and 2MASS infrared
$JHK_s$ data, in combination with optical photometry. Except for two
clusters, the ages of the other sample GCs are all older than 1
Gyr. Their age distribution shows that most sample clusters are
younger than 6 Gyr, with a peak at $\sim 3$ Gyr, although the `usual'
complement of well-known old GCs (i.e., GCs of similar age as the
majority of the Galactic GCs) is present as well.
\end{abstract}

\keywords{galaxies: individual (M31) -- galaxies: star clusters --
galaxies: stellar content}

\section{Introduction}
\label{Introduction.sec}

The process of galaxy formation and evolution ranks among the most
important outstanding problems in astrophysics
\citep[e.g.,][]{per02}. One way to better understand the underlying
questions is by studying globular clusters (GCs). GCs are often
considered the fossils of the galaxy formation process, since they
tend to form in the very early stage of their host galaxy's evolution
\citep{bh00}. A GC is a densely packed, gravitationally bound, roughly
spherical system of several thousand to about one million stars. They
can be observed out to great distances, which implies that they can be
used to study and probe the properties of extragalactic systems. The
most distant GC systems studied to date are located in the Coma
Cluster; their study has been facilitated by {\sl Hubble Space
Telescope} ({\sl HST}) Wide Field and Planetary Camera-2 (WFPC2)
observations \citep{baum95,Kavelaars00,harris00,wh00}.

M31 (NGC 224), the Andromeda galaxy, is an early-type spiral galaxy
(type Sb), located at a distance of $\sim 780$ kpc \citep{sg98,
mac01}. It is the nearest and largest spiral galaxy in the Local
Group of galaxies and has been the subject of many GC studies and
surveys. \citet{hubble32} first discovered 140 GC candidates
characterized by $m_{\rm pg}\leq 18$ mag. Subsequently, a number of
studies \citep{seynas45,hilt58,mayegg53,km60} identified $\sim 160$
GC candidates in M31. \citet{vete62a} compiled the first major M31
GC catalog, containing about 300 GC candidates and including $UBV$
photometric data. The most extensive GC surveys have since been
published by \citet{sarg77}, \citet{Crampton85}, and the Bologna
group \citep{batt80,batt87,batt93}. In particular,
\citet{Crampton85} and \citet{batt80,batt87,batt93} provided
photometric data in either $UBV$ or $UBVR$. These surveys were
mostly based on visual searches of photographic plates and are
fairly complete down to $V=18$ mag ($M_V \sim -6.5$ mag)
\citep{fusi1993}, although a number of recent studies searched for
fainter GCs in M31 \citep[e.g.,] []{moche98,bh01,kim07}. However,
\citet{gall06} showed that a significant number of class-D and E GCs
with $V\ga 17$ are still to be confirmed (and hence the GC
luminosity function is incomplete), and that large surveys are
needed to reach a complete sample of M31 GCs. Following on from the
first extensive spectroscopic survey of M31 GCs by \citet{van69}, a
significant number of authors \citep[e.g.,] [and references
therein]{hsv82,hbk91,DG,luciana,jab,bh00,per02,gall06,lee08}
embarked on studies of their spatial, kinematic, and metal-abundance
properties. The first comprehensive catalog including photometric
and spectroscopic data for M31 GCs was assembled by \citet{bh00}.
The Revised Bologna Catalog (RBC) of M31 GCs was recently published
by \citet{gall04} and has since been revised a number of times
\citep{gall05,gall06,gall07}. In the primary catalog \citep{gall04},
all known M31 GCs and candidates were compiled based on a literature
survey, leading to a total of 1164 entries including 337 confirmed
GCs, 688 GC candidates, and 10 objects with undetermined
classification. In addition, \citet{gall04} identified 693 known and
candidate GCs in M31 using the 2MASS database and included their
2MASS $JHK_s$ magnitudes. The latest RBC (V3.5) was updated on March
27, 2008, and includes the newly discovered star clusters from
\citet{mac06}, \citet{kim07}, and \citet{hux08}. In total, 1567 GCs
and GC candidates (509 confirmed GCs and 1058 GC candidates) are
known in M31; 421 former GC candidates turned out to be stars,
asterisms, galaxies, H{\sc ii} regions, or extended clusters. In
addition, the RBC V3.5 includes photometric data of M31 GCs and GC
candidates in the far- and near-ultraviolet (FUV and NUV) from the
Nearby Galaxies Survey (NGS) of the {\sl Galaxy Evolution Explorer}
({\sl GALEX}) \citep{rey06}. Very recently, \citet{caldwell08}
presented a new catalog of 670 likely star clusters in the field of
M31, all with updated high-quality coordinates accurate to $0.2''$,
based on images from either the Local Group Survey \citep{massey06}
or the Digital Sky Survey.

An accurate and reliable analysis of integrated stellar populations
(such as star clusters) is key to understanding the formation and
evolutionary process in galaxies. By means of comparisons of
integrated populations with models of homogeneous stellar systems,
i.e., simple stellar populations (SSPs), recent studies have achieved
some success in determining ages and masses for extragalactic star
clusters \citep[e.g.,]
[]{degrijs03a,degrijs03b,degrijs03c,bik03,ma06,fan06,ma07a}.

\citet{ma06} and \citet{fan06} derived age estimates for M31 GCs by
fitting SSP models \citep[][henceforth BC03]{bru03} to their
photometric measurements in a large number of intermediate- and
broad-band passbands from the optical to the near-infrared (NIR).
For instance, \citet{ma07a} constrained the age of the M31 GC S312
(B379), using multicolor photometry from the NUV to the NIR, to
$9.5^{+1.15}_{-0.99}$~Gyr. S312 (B379) is among the first
extragalactic GCs for which the age was estimated accurately using
main-sequence photometry, i.e., \citet{brown04} estimated its age at
$10^{+2.5}_{-1}$ Gyr. This was based on their analysis of the
cluster's color-magnitude diagram (CMD) below the main-sequence
turn-off (MSTO) using extremely deep images obtained with the
Advanced Camera for Surveys (ACS) on board the {\sl HST}. They
performed a quantitative comparison of their resolved stellar
photometry with the isochrones of \citet{vandenberg06}.

In this paper we first describe our new observations and the relevant
data-processing steps, as well as the complementary data used from the
literature (\S\ref{data.sec}). In \S\ref{fit.sec} we quantitatively
compare the spectral energy distributions (SEDs) of the GCs in our
sample with the {\sc galev} SSP models. Finally, our results and a
summary are presented in \S\ref{result.sec}.

\section{Database}
\label{data.sec}

\subsection{The sample}

The GC sample used in this paper was taken from the Bologna catalog
of \citet{batt87}, which contains 827 M31 GCs and GC candidates. In
addition, our sample also supplements that of \citet{jiang03}, who
studied 172 GC candidates from \citet{batt87} on the basis of
integrated photometric measurements in the
Beijing-Arizona-Taiwan-Connecticut (BATC) photometric system. In
\citet{jiang03}, all GC candidates of classes A and B (353 objects)
in \citet{batt87} (i.e., their Table IV) were adopted as their
original sample. However, of these only 223 objects are in their
observed CCD fields. They also noted that B007 is a galaxy, and
B055, B132 and B147 are virtually stars \citep{bh00}. These four
objects were therefore not included in \citet{jiang03}'s final
sample. In summary, 219 class-A or B GCs were observed by
\citet{jiang03}, of which 47 were excluded because of missing
photometric measurements in some filters. In this paper we analyze
these 47 GC candidates on the basis of newly observed data in the
BATC photometric system combined with {\sl GALEX} FUV/NUV
photometry, broad-band $UBVRI$, and NIR $JHK_s$ (2MASS) data.
However, we did not manage to obtain accurate photometric
measurements for a number of objects because of either the dominance
of a nearby very bright object (B095, B176, and B202), the GC
candidate being very faint and superimposed on a very high
background (B119 and B324), or the GC candidate being located near
M32 (B124) or NGC 205 (B331), both also resulting in a very high
contribution. In addition, object B330 is faint and located very
close to a brighter object, rendering accurate photometry
impossible. Thus, here we analyze the multicolor photometric
properties of 39 GC candidates. Figure 1 shows the spatial
distributions of both our sample GCs (circles) and the
\citet{jiang03} GCs (plus signs) across the M31 field.

\begin{figure}
\resizebox{\hsize}{!}{\rotatebox{0}{\includegraphics{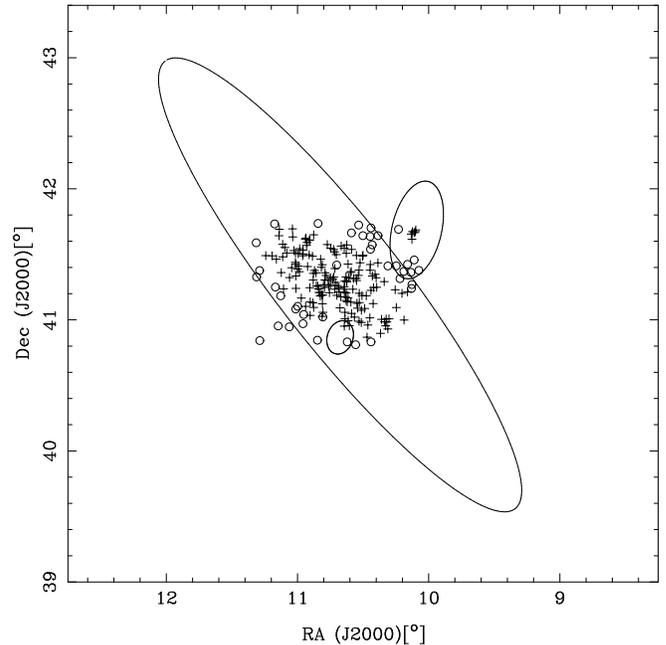}}}
\caption{Spatial distribution of the GC candidates in M31. Circles and
plus signs represent the samples discussed in this paper and by
\citet{jiang03}, respectively. The large ellipse is the M31 disk/halo
boundary as defined by \citet{rac91}; the two small ellipses are the
$D_{25}$ isophotes of NGC 205 (northwest) and M32 (southeast).}
\label{fig1}
\end{figure}

\subsection{Observations and data reduction}

To obtain photometric measurements in the BATC photometric system
for the 39 GC candidates for which \citet{jiang03} did not obtain
photometry in a number of filters, we re-observed the objects. The
BATC photometric system uses a Ford Aerospace 2048$\times$2048 CCD
camera with a pixel size of 15 $\mu$m, mounted at the focus of the
0.6/0.9m $f$/3 Schmidt telescope at Xinglong Station (National
Astronomical Observatories of China; NAOC). The CCD field of view is
$58^{\prime}\times 58^{\prime}$, with a pixel size of $1.7''$. The
multicolor BATC filter system includes 15 intermediate-band filters
covering the wavelength range from 3300{\AA} to 1 $\mu$m. These
filters were specifically designed to avoid contamination from the
brightest and most variable night-sky emission lines. The CCD camera
is not sensitive at the shortest wavelengths covered by the BATC
filters. For this reason, neither \citet{jiang03} nor we used the
two bluest filters ($a$ and $b$) for our observations. Finding
charts of the sample GCs and GC candidates in the BATC $g$ band
(centered at 5795\AA), obtained with the NAOC 60/90cm Schmidt
telescope, are shown in Fig. 2.

Thirteen hours of imaging of the M31 field of \citet{jiang03} were
obtained through the usable set of 13 intermediate-band filters from
November 15, 2003 to December 13, 2003. Bias subtraction and flat
fielding using dome flats were done with the BATC automatic
data-reduction software, {\sc pipeline i}, originally developed for
the BATC Multicolor Sky Survey \citep{fan96,zheng99}. The dome
flat-field images were taken using a diffuser plate in front of the
Schmidt telescope's corrector plate. This flat-fielding technique was
verified using photometry obtained for other galaxies and
spectrophotometric observations \citep[see, e.g.,]
[]{fan96,zheng99,wu02,yan00,zhou01,zhou04}. Spectrophotometric
calibration of the M31 images was done by observations of four F-type
subdwarfs, HD~19445, HD~84937, BD~${+26^{\circ}2606}$, and
BD~${+17^{\circ}4708}$ \citep{ok83}. Our magnitudes are therefore
defined in the spectrophotometric AB magnitude system (i.e., the Oke
\& Gunn $\tilde{f_{\nu}}$ monochromatic system),

\begin{equation}
m_{\rm BATC}=-2.5{\rm log}\tilde{F_{\nu}}-48.60,
\end{equation}
where $\tilde{F_{\nu}}$ is the appropriately averaged monochromatic
flux in units of erg s$^{-1}$ cm$^{-2}$ Hz$^{-1}$ at the effective
wavelength of the specific passband. In the BATC system
$\tilde{F_{\nu}}$ is defined as \citep{yan00}

\begin{equation}
\tilde{F_{\nu}}=\frac{\int{\rm d} ({\rm log}\nu)f_{\nu}r_{\nu}}
{\int{\rm d} ({\rm log}\nu)r_{\nu}},
\end{equation}
which relates the magnitude to the number of photons detected rather
than to the input flux \citep{fuku96}. In Eq. (2), $r_{\nu}$ is the
system's response and $f_{\nu}$ the object's SED.  Spectrophotometric
calibration of the M31 images using the Oke-Gunn standard stars was
done during photometric nights \citep[see for details]
[]{yan00,zhou01}. Using these standard-star images, we iteratively
obtained atmospheric extinction curves and the variation of these
extinction coefficients as a function of the time of night \citep[cf.]
[]{yan00,zhou01},

\begin{equation}
m_{\rm BATC}=m_{\rm inst}+[K+\Delta K({\rm UT})]X+C,
\end{equation}
where $X$ is the airmass and $[K+\Delta K({\rm UT})]$ the
time-dependent extinction term. The instrumental magnitudes ($m_{\rm
inst}$) of selected bright, isolated, and unsaturated stars on the M31
images observed on photometric nights can be readily transformed to
the BATC system ($m_{\rm BATC}$). The calibrated magnitudes of these
stars were then used as secondary standards to uniformly combine
images from calibrated nights with their counterparts observed on
non-photometric nights. Table 1 lists the parameters of the BATC
filters and the observational statistics; column 6 provides the
scatter, in magnitudes, for the photometric observations of the four
primary standard stars in each filter.

\begin{figure*}
\resizebox{\hsize}{!}{\rotatebox{-0}{\includegraphics{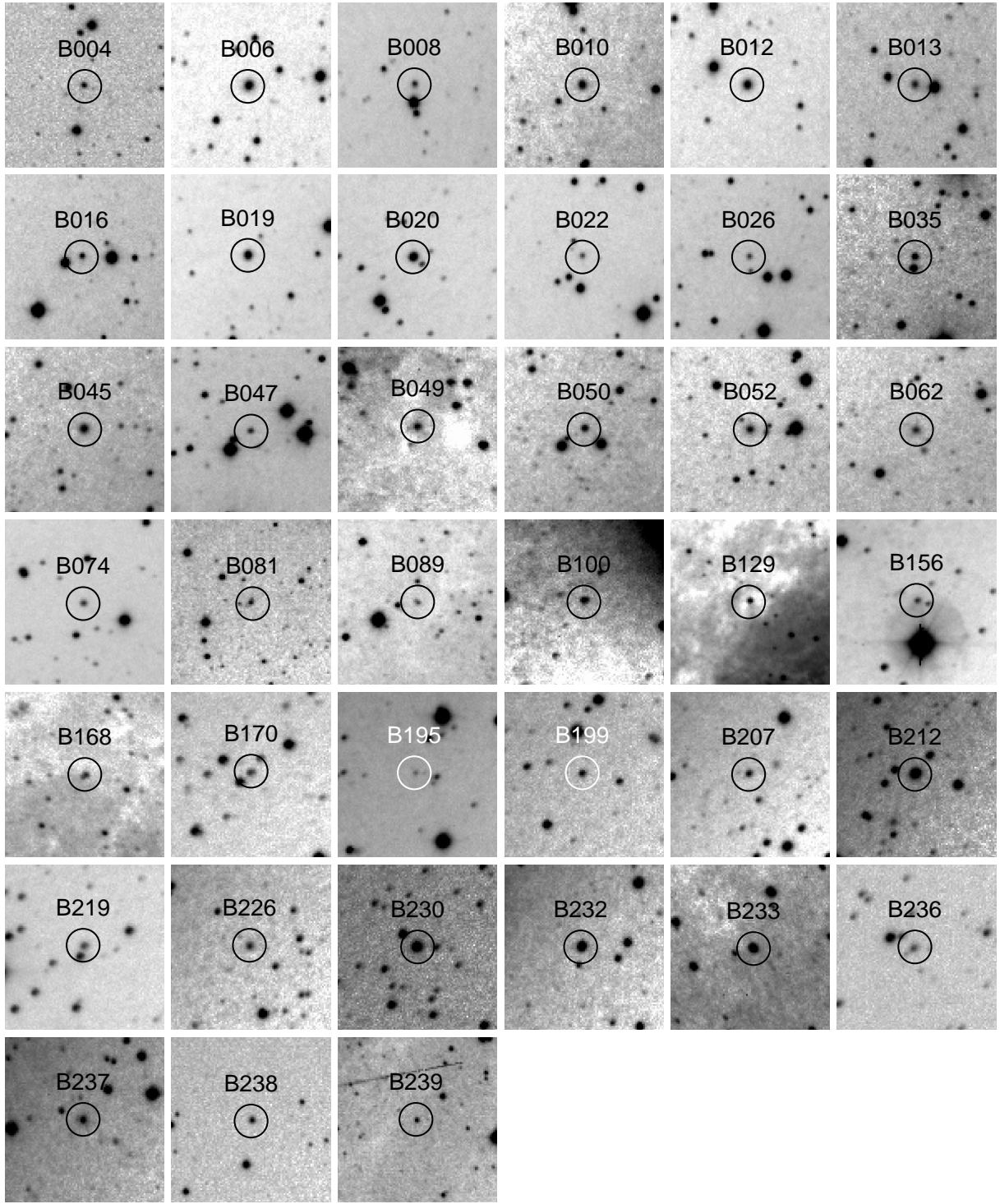}}}
\vspace{-4.5cm} \caption{Finding charts of the sample GCs and GC
candidates in the BATC $g$ band, obtained with the NAOC 60/90cm
Schmidt telescope. The field of view of each image is
$11^{\prime}\times 11^{\prime}$.} \label{fig2}
\end{figure*}

\subsection{Integrated photometry}

For each M31 GC candidate we used the {\sc phot} routine in DAOPHOT
\citep{stet87} to obtain the integrated photometry. To avoid
contamination from nearby objects, we adopted an aperture of
$10.2''$ diameter, corresponding to 6 pixels. Inner and outer radii
for background determination were taken at 8 and 13 pixels from the
GC center. Given the small aperture used for the GC observations,
aperture corrections were determined as follows. We used isolated
stars to determine the magnitude difference between diameter of 6
pixels and the fully integrated magnitude of these stars in each of
the 13 BATC filters used. The SEDs for our sample of 39 GCs and GC
candidates were then corrected for the filter-specific differences,
and these values are given in Table 2. Columns 2--14 give the
magnitudes in the 13 BATC passbands used for our observations. For
each object the second line lists the $1\sigma$ uncertainties in
magnitude for the corresponding passband. The errors for each filter
are given by DAOPHOT. The magnitudes of B129 in the $c$ and $d$
filters, and that of B195 in $p$ filter could not be obtained owing
to low signal-to-noise ratios in these filters.

\subsection{GALEX, broad-band, and 2MASS photometry}

To estimate the ages of the M31 sample GCs and GC candidates
accurately, we use as many photometric data points covering as wide a
wavelength range as possible \citep[cf.][]{degrijs03b,Anders04}. In
addition, \citet{kaviraj07} showed that the combination of FUV and NUV
photometry with optical observations in the standard broad bands
enables one to efficiently break the age-metallicity
degeneracy. \citet{wor94} showed that the age-metallicity degeneracy
associated with optical broad-band colors is $\Delta {\rm age}/\Delta
Z \sim 3/2$ \citep[see also][]{MacArthur04}. However, \citet{jong96}
showed that this degeneracy can be partially broken by adding NIR
photometry to optical colors, which was recently confirmed by
\citet{wu05}. \citet{Cardiel03} found that the inclusion of an
infrared (IR) passband can improve the predictive power of the stellar
population diagnostics by $\sim 30$ times compared to using optical
photometry alone. Since NIR photometry is less sensitive to
interstellar extinction than the classical optical passbands,
\citet{kbm02} and \citet{puzia02a} also suggested that it provides
useful complementary information that can help to disentangle the
age-metallicity degeneracy \citep[also see][]{gall04}.

\citet{rey06} published {\sl GALEX} NUV and FUV photometric data for
485 and 273 M31 GCs, respectively. The photometric data for 28 (NUV)
and 17 (FUV) of our M31 sample GCs in common is listed in Table
3. Again, for each object the second line lists the photometric
uncertainties for the corresponding passband. The {\sl GALEX}
photometric system is calibrated to match the spectrophotometric AB
system.

To date, the study of M31 GCs has been largely based on the excellent
Bologna catalog \citep{batt80,batt87,batt93}. Updates to the original
RBC were provided by \citet{gall04} who take as their photometric
reference the dataset of \citet{bh00} in order to obtain the most
homogeneous set of photometric measurements available. \citet{bh00}
published optical and IR photometric data for 285 M31 GCs (see their
Table 3), obtained with the 4-Shooter CCD mosaic camera and the SAO IR
imager on the 1.2m telescope at the Fred Lawrence Whipple
Observatory. Photometric measurements in the $UBVRI$ bands were
published by \citet{bh00} for most of our sample objects. Therefore,
we preferentially adopt the $UBVRI$ measurements of \citet{bh00}. For
the remaining GCs we follow \citet{gall04}, who updated the Bologna
catalog with homogenised optical ($UBVRI$) photometry collected from
the most recent photometric references available.  \citet{gall04} did
not include the photometric uncertainties.  Although we refer to the
original papers, the uncertainties associated with the same object but
based on the use of different photometric systems are often very
different. In addition, \citet{gall04} transformed their $UBVRI$
photometry to the reference system of \citet{bh00} by applying offsets
derived from objects in common between the relevant catalog and the
data set of \citet{bh00}. The measurements are therefore internally
consistent, and referencing the original uncertainties may be
irrelevant. Therefore, we only adopted photometric uncertainties as
suggested by \citet{gall04}, i.e., 0.05 mag in $BVRI$ and 0.08 mag in
$U$. In fact, these photometric uncertainties do not affect our
results significantly, as we showed in \citet{fan06} (see their \S4.3
for details).

\citet{gall04} identified 693 known and candidate GCs in M31 using the
2MASS database, and determined their 2MASS $JHK_{s}$ photometric
magnitudes (transformed to the CIT photometric system) \citep{Elias82,
Elias83}. However, we need the original 2MASS $JHK_{s}$ magnitudes for
our sample GCs to compare our observational SEDs with the SSP models,
so we reversed this transformation using the same procedures. Since
\citet{gall04} did not provide the 2MASS $JHK_s$ uncertainties, we
obtained these by comparing the photometric magnitudes with Fig. 2 of
\citet{Carpenteretal01}. They show the observed photometric rms
uncertainties as a function of magnitude for stars brighter than their
observational completeness limits. We include the broad-band and 2MASS
photometry (and the associated uncertainties) of the sample GCs in
Table 3. We also list the new classification flags, following RBC V3.5
notation. From Table 3 we learn that B052 and B062 are classified as
galaxies based on their radial velocities. We will therefore not
estimate their ages below.

\subsection{Comparison with previously published photometry}

The BATC intermediate-band system can easily be transformed to the
$UBVRI$ broad-band system. \citet{zhou03} derived the
relationships between these two systems using standard stars from the
catalogs of \citet{lan83,lan92} and \citet{gala00}:

\begin{equation}
m_B=m_{d}+0.2201(m_{c}-m_{e})+0.1278\pm0.076,
\end{equation}

\begin{equation}
m_V=m_{g}+0.3292(m_{f}-m_{h})+0.0476\pm0.027.
\end{equation}

To check our photometry we derived the magnitudes in $B$ and $V$ based
on Eqs. (4) and (5). We transformed the magnitudes of our 39 GCs and
GC candidates in the BATC $c, d$, $e$ bands to $B$-band photometry,
and BATC $f, g$, and $h$-band measurements into $V$-band data. Fig. 3
shows a comparison of our $V$ and $(B-V)$ photometry with previously
published measurements of \citet{bh00} and \citet{gall04}. The mean
$V$ magnitude and $(B-V)$ color differences (in the sense of this
paper minus \citet{bh00} or \citet{gall04}) are $\langle \Delta V
\rangle =-0.066\pm0.013$ mag and $\langle \Delta (B-V) \rangle
=-0.040\pm 0.017$ mag, respectively, thus showing excellent agreement.

\begin{figure}
\resizebox{\hsize}{!}{\rotatebox{-90}{\includegraphics{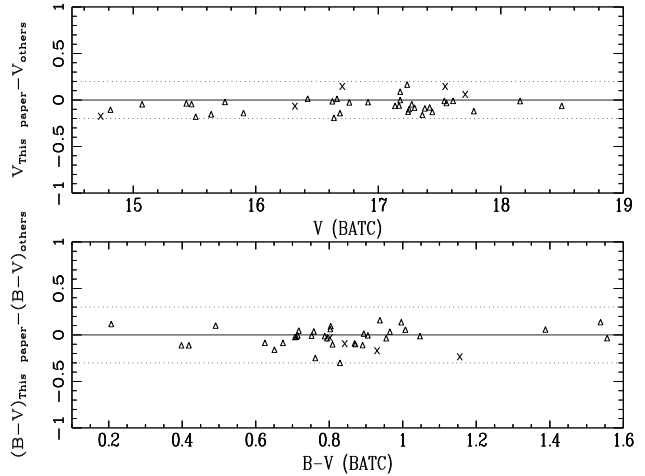}}}
\caption{Comparison of our newly obtained cluster photometry with
previous measurements by \citet{bh00} (triangles) and \citet{gall04}
(crosses). The dashed lines enclose $\pm 0.2$ mag in $V$ and $\pm 0.3$
mag in $B-V$.} \label{fig3}
\end{figure}

\subsection{Metallicities and reddening values}

To estimate the ages of our sample GCs accurately we required that
our GCs have both independently determined metallicities and
reddening values. We used three homogeneous sources for
spectroscopic metallicities \citep{hbk91,bh00,per02} and one
reference \citep{fan08}.

\citet{hbk91} obtained spectroscopy of 150 M31 GCs and candidates with
the Multiple Mirror Telescope (MMT). The system they used has a
resolution of 8--9{\AA} and enhanced blue sensitivity. To obtain many
of the strongest and most metallicity-sensitive spectral features of
interest in the ultraviolet, they extended their observations to the
atmospheric cut-off at 3200{\AA} \citep[see details in] []{bh90}. The
metallicities of these 150 objects were determined using six
absorption-line indices from integrated cluster spectra employing the
method of \citet{bh90}.

\citet{bh00} observed 61 M31 GCs and candidates spectroscopically
using the Keck Low Resolution Imaging Spectrometer (LRIS) and the MMT
Blue Channel spectrograph. With Keck LRIS, they used a 600 $\ell$
mm$^{-1}$ grating with a 1.2{\AA} pixel$^{-1}$ dispersion from
3670--6200{\AA}, and a resolution of 4--5\AA. With the MMT Blue
Channel, they used a 300 $\ell$ mm$^{-1}$ grating with a 3.2{\AA}
pixel$^{-1}$ dispersion from 3400--7200 {\AA}, and a resolution of
9--11{\AA}. They obtained the cluster metallicities on the basis of
the \citet{bh90} method as well.

\citet{per02} determined metallicities for more than 200 M31 GCs and
candidates using the Wide Field Fibre Optic Spectrograph (WYFFOS) at
the 4.2m William Herschel Telescope. Their spectral range covers
$\sim$ 3700--5600{\AA} using two gratings, one of which (H2400B, 2400
$\ell$ mm$^{-1}$) yields a dispersion of 0.8{\AA} pixel$^{-1}$ and a
spectral resolution of 2.5{\AA} over the range 3700--4500{\AA}, and
the other (R1200R, 1200 $\ell$ mm$^{-1}$) yields a dispersion of
1.5{\AA} pixel$^{-1}$ and a spectral resolution of 5.1{\AA} over the
range 4400--5600{\AA}. \citet{per02} calculated 12 absorption-line
indices, again using the method of \citet{bh90}. Through a comparison
of the line indices with published M31 GC [Fe/H] values from previous
studies \citep{bonoli87,bh90,bh00}, they found that the line indices
of the CH (G band), Mg$b$, and Fe53 lines best represented their
observed GCs. Therefore, \citet{per02} determined the metallicities of
their sample targets using an unweighted mean of these three [Fe/H]
values.

Using metallicities from the literature \citep{hbk91,bh00,per02}
combined with the RBC, \citet{fan08} determined 443 reddening values
and intrinsic colors, as well as 209 metallicities for individual
clusters without spectroscopic observations.

To use all metallicities as coherently as possible we ranked the
sources of M31 GCs metallicities, choosing \citet{per02} metallicities
whenever available because of the large number of metallicity
determinations. Metallicities from \citet{bh00} and \citet{hbk91} were
preferred if \citet{per02} determinations were not available. If
spectroscopic metallicities were missing, we used
\citet{fan08}. Metallicities were not available for B089 and B226. As
a consequence, we do not attempt to determine their ages (see details
in \S4). The final set of metallicities for the sample clusters is
included in Table 4.

For the reddening values of the sample GCs we refer to \citet{bh00}
and \citet{fan08}. \citet{bh00} determined the reddening for each
cluster using correlations between optical and IR colors and
metallicity, and by defining various `reddening-free' parameters
using their large database of multicolor photometry.  \citet{bh00}
found that the M31 and Galactic GC extinction laws, and the M31 and
Galactic GC color-metallicity relations are similar. They estimated
the reddening to M31 objects with spectroscopic data using the
relationship between intrinsic optical color and metallicity for
Galactic clusters. For objects without spectroscopic data they used
the relationships between the reddening-free parameters and certain
intrinsic colors based on Galactic GC data. Following the methods in
\citet{bh00}, \citet{fan08} (re-)determined reddening values for 443
clusters and cluster candidates. We choose \citet{fan08} reddening
values whenever available because their reddening values comprise a
homogeneous data set and they are larger than those of \citet{bh00}.
The reddening values for the sample clusters are listed in Table 4.
The values of extinction coefficient $R_{\lambda}$ are obtained by
interpolating the interstellar extinction curve of \citet{car89}.

\section{Age determination}
\label{fit.sec}

\subsection{Stellar populations and synthetic photometry}

The most direct method to constrain the ages of different stellar
populations involves comparing the observed luminosity levels of the
MSTOs. Unfortunately, this approach is limited to the nearest GCs,
where individual stars can be resolved and measured down to a few
magnitudes fainter than the MSTO. Even in M31, the nearest large
spiral galaxy, the MSTO is only reached for one GC (S312) \citep[also
see][]{brown04,rey06,ma07a}. However, since the pioneering work of
\citet{Tinsley68,Tinsley72} and \citet{ssb73} evolutionary population
synthesis modeling has become a powerful tool for the interpretation
of integrated spectrophotometric observations of galaxies as well as
their components \citep[see e.g.][]{Anders04}.

In evolutionary synthesis models, SSPs are modeled on the basis of a
collection of evolutionary tracks of stars of different initial masses
and a set of stellar spectra at different evolutionary stages. To
estimate the ages of our sample GCs we compare their SEDs with the
{\sc galev} SSP models \citep[e.g.,][]{kurth99,schulz02,anders03}. The
{\sc galev} SSPs are based on the Padova isochrones (with the most
recent versions using the updated \citet{bertelli04} isochrones, which
include the thermally-pulsing asymptotic giant-branch [TP-AGB] phase),
and a \citet{sal55} stellar initial mass function with a lower-mass
limit of $0.10~{\rm M}_\odot$ and the upper-mass limit between 50 and
70 ${\rm M}_\odot$, depending on metallicity. The full set of models
spans the wavelength range from 91{\AA} to 160 $\mu$m. These models
cover ages from $4 \times 10^6$ to $1.6 \times 10^{10}$ yr, with an
age resolution of 4 Myr for ages up to 2.35 Gyr, and 20 Myr for
greater ages.

Since our observational data consists of integrated luminosities
through a given set of filters, we convolved the theoretical SSP SEDs
with the {\sl GALEX} FUV and NUV, broad-band $UBVRI$, BATC, and 2MASS
$JHK_s$ filter response curves to obtain synthetic ultraviolet,
optical, and NIR photometry for comparison. The synthetic magnitude in
the AB magnitude system for the $i^{\rm th}$ filter can be computed as

\begin{equation}
m_i=-2.5\log\frac{\int_{\lambda}F_{\lambda}\varphi_{i} (\lambda){\rm
d}\lambda}{\int_{\lambda}\varphi_{i}(\lambda){\rm d}\lambda}-48.60,
\end{equation}
where $F_{\lambda}$ is the theoretical SED and $\varphi_i$ the
response curve of the $i^{\rm th}$ filter of the {\sl GALEX} FUV/NUV,
$UBVRI$, BATC, and 2MASS $JHK_{s}$ photometric systems. Here,
$F_{\lambda}$ changes as a function of age and metallicity.

\subsection{Fitting results}

We use a $\chi^2$ minimization test to examine which {\sc galev} SSP
models are most compatible with the observed SEDs, following
\begin{equation}
\chi^2=\sum_{i=1}^{23}{\frac{[m_{\lambda_i}^{\rm
intr}-m_{\lambda_i}^{\rm mod}(t)]^2}{\sigma_{i}^{2}}},
\end{equation}
where $m_{\lambda_i}^{\rm mod}(t)$ is the integrated magnitude in
the $i^{\rm th}$ filter of a theoretical SSP at age $t$,
$m_{\lambda_i}^{\rm intr}$ represents the intrinsic integrated
magnitude in the same filter, and
\begin{equation}
\sigma_i^{2}=\sigma_{{\rm obs},i}^{2}+\sigma_{{\rm mod},i}^{2}.
\end{equation}

Here, $\sigma_{{\rm obs},i}^{2}$ is the observational uncertainty, and
$\sigma_{{\rm mod},i}^{2}$ is the uncertainty associated with the
model itself, for the $i^{\rm th}$ filter. \citet{charlot96} estimated
the uncertainty associated with the term $\sigma_{{\rm mod},i}^{2}$ by
comparing the colors obtained from different stellar evolutionary
tracks and spectral libraries. Following \citet{wu05}, \citet{ma06},
and \citet{fan06} we adopt $\sigma_{{\rm mod},i}^{2}=0.05$. In fact,
the values $\sigma_{{\rm mod},i}^{2}$ adopted do not change the best
fits, but only affect the $\chi^2$ values.

The {\sc galev} SSP models include five initial metallicities,
$Z=0.0004, 0.004, 0.008, 0.02$ (solar metallicity), and 0.05. Spectra
for other metallicities can be obtained by linear interpolation of the
appropriate spectra for any of these metallicities. In addition, if
the metallicity of a cluster is poorer than $Z=0.0004$, we only use
the model of $Z=0.0004$. The best fits to the SEDs of our GCs are
presented in Fig. 4.

\begin{figure*}
\figurenum{4}
\resizebox{\hsize}{!}{\rotatebox{-0}{\includegraphics{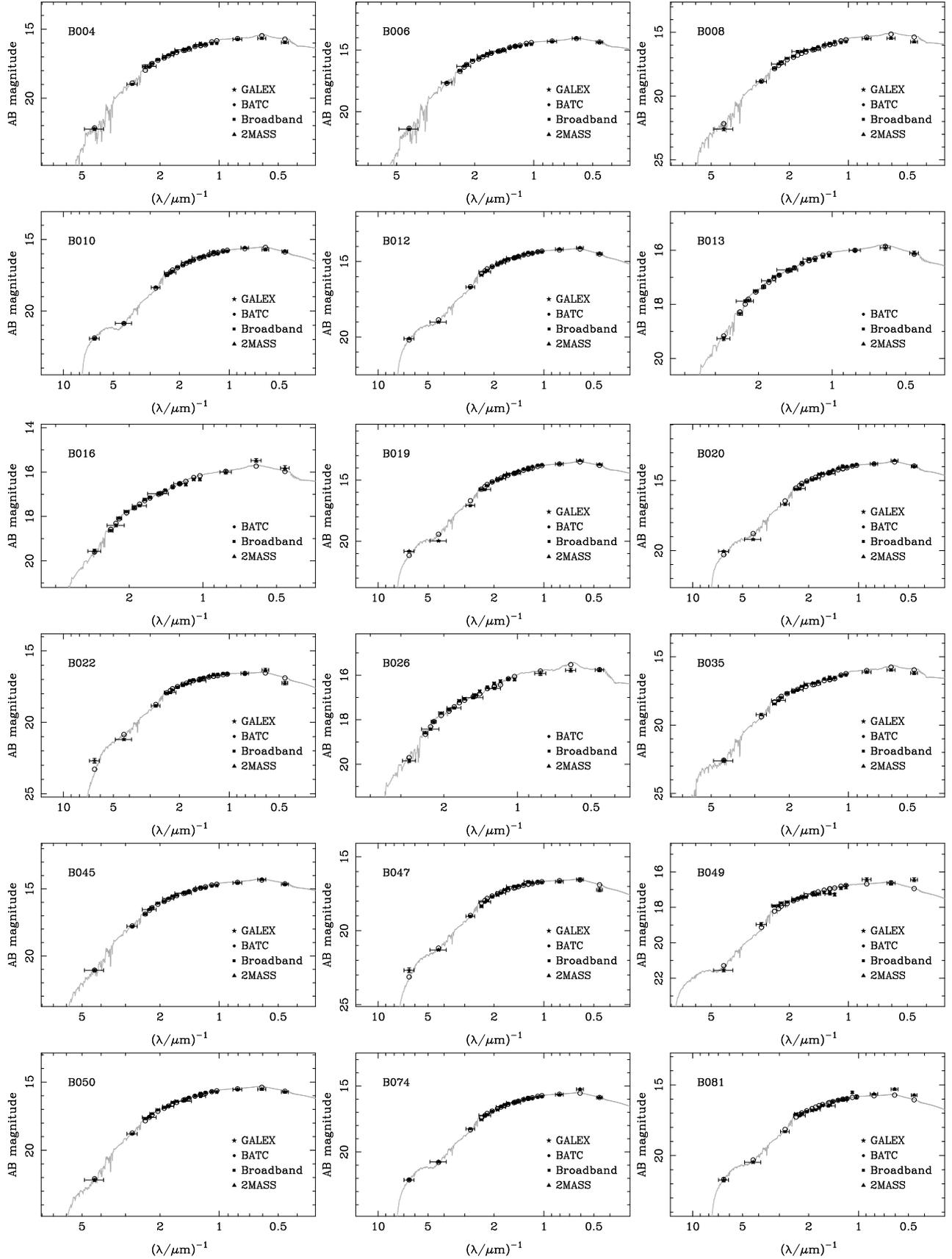}}}
\vspace{-1.5cm} \caption{Best-fitting integrated SEDs of the {\sc
galev} SSP models shown in relation to the intrinsic SEDs for our
sample GCs. The photometric data points are represented by the symbols
with error bars (vertical error bars for uncertainties and horizontal
ones for the approximate wavelength coverage of each filter). Open
circles represent the calculated magnitude of the model SEDs for each
filter.} \label{fig4}
\end{figure*}

\begin{figure*}
\figurenum{4}
\resizebox{\hsize}{!}{\rotatebox{-0}{\includegraphics{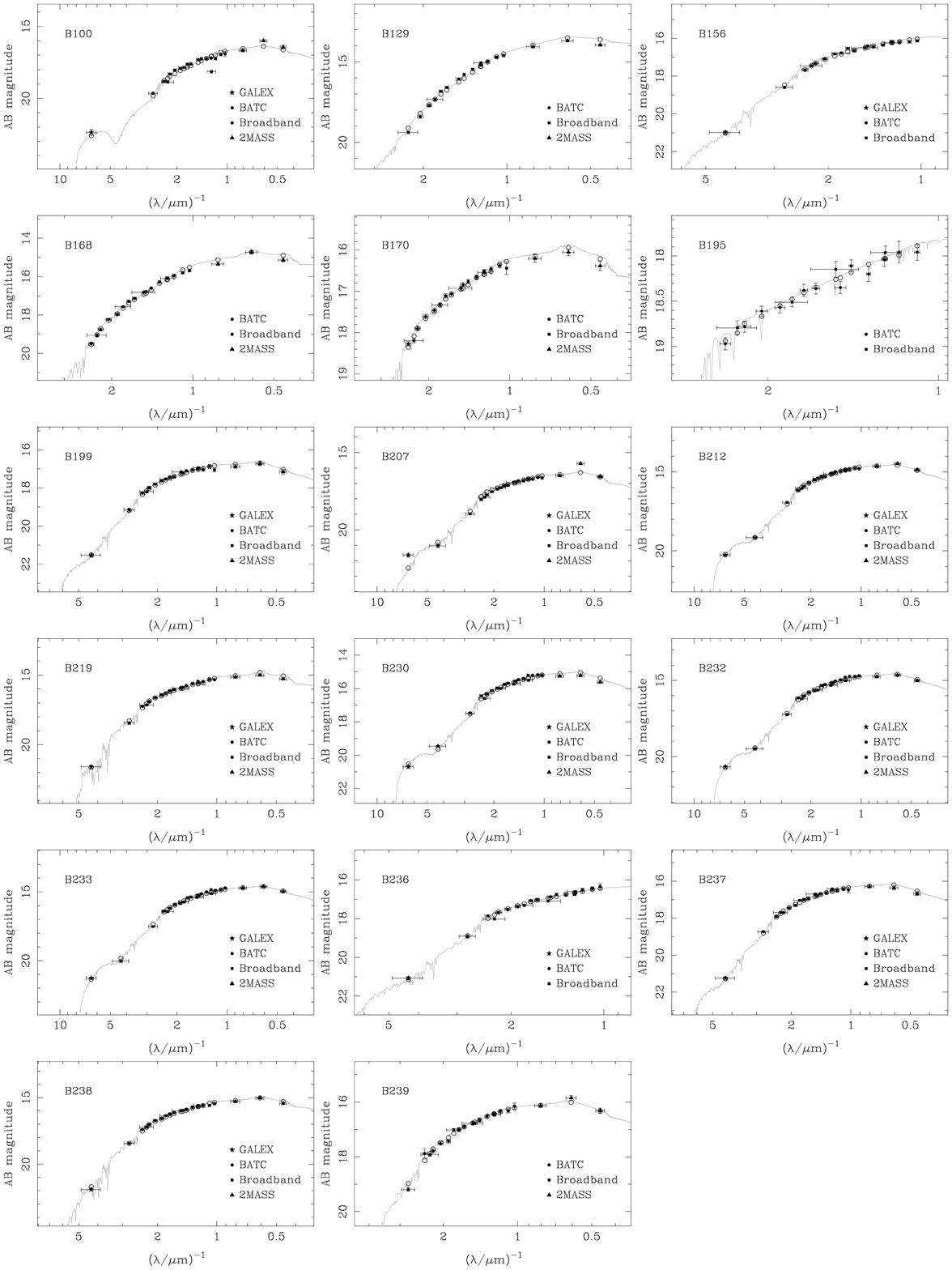}}}
\vspace{-1.5cm} \caption{Continued.} \label{fig4}
\end{figure*}

\section{Results and summary}
\label{result.sec}

In the previous Section we determined the ages of 35 M31 GCs and GC
candidates. The results are listed in Table 5. The metallicity of B089
and B226, and the reddening of B089 had not been determined
previously. From Fig. 5, which shows the age distribution of the
sample clusters (see also Table 5) we conclude that, except for two
clusters, the ages of the other sample GCs are all older than 1
Gyr. Most sample GCs are younger than 6 Gyr, with a peak at $\sim 3$
Gyr. The `usual' complement of well-known old GCs (i.e., GCs of
similar age as the majority of the Galactic GCs) is also present.

\begin{figure}
\figurenum{5}
\resizebox{\hsize}{!}{\rotatebox{-90}{\includegraphics{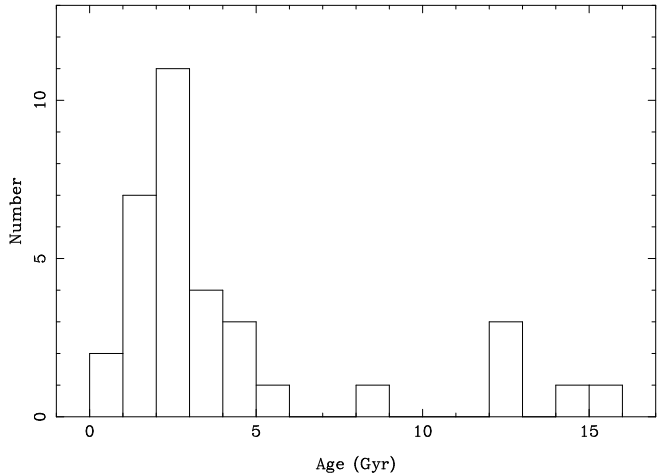}}}
\caption{Age distribution of our sample GCs and GC candidates in
M31.} \label{fig5}
\end{figure}

As discussed in \S 2.6, to estimate the ages of our sample GCs
accurately we required that our GC sample have both independently
determined metallicities and reddening values. For metallicity, we
used \citet{hbk91}, \citet{bh00}, and \citet{per02} as our reference
data set. Since all of these authors determined the M31 GC
metallicities using the calibration of \citet{bh90}, all three
metallicity determinations are on the same [Fe/H] scale and there
are no systematic offsets among any of these data sets \citep[see
details in][]{per02}. However, individual metallicity differences
exist, which may affect our age estimates. Twelve GCs and GC
candidates have two metallicity determinations, of which B004, B219,
and B238 exhibit the largest differences ($> 0.2$ dex). The
metallicities of B004, B219, and B238 from \citet{per02} are
$-0.31\pm0.74$, $-0.01\pm 0.57$, and $-0.57\pm0.66$ dex, compared to
$-1.26\pm 0.59$, $-0.53\pm0.53$, and $-1.22\pm0.76$ dex from
\citet{hbk91}. Large metallicity differences lead to large age
differences. The ages of B004, B219, and B238 are estimated at
$4.10\pm0.55$, $2.50\pm0.15$, and $5.00\pm0.45$ Gyr \citep[based on
the metallicities of ][]{per02}, compared to $14.40\pm0.75$,
$11.60\pm1.45$ and $14.40\pm0.80$ Gyr \citep[metallicities
of][]{hbk91}. This implies that the accuracy of the metallicity
determinations is very important for the corresponding age
estimates. The age differences for the other 10 GCs are less than 1
Gyr based on the metallicities of both \citet{per02} and
\citet{hbk91}, except for B045: $8.80\pm1.45$ Gyr versus
$6.30\pm0.45$ Gyr based on \citet{per02} and \citet{hbk91}
metallicities, respectively. In general, the three different sources
of spectroscopic metallicities provide homogeneous age estimates.
For B004, B219, and B238 the signal-to-noise ratios of the
observations of \citet{hbk91} and \citet{per02} are too low.
High-quality spectral observations of these three GCs are needed. In
addition, we point out that the metallicity calibration of
\citet{bh90} is solely based on old GCs. However, the sample GCs and
GC candidates discussed in this paper are estimated to be young or
of intermediate age, so that the age estimates may be somewhat
biased by the adopted calibration.

\citet{bh00} discovered that M31 contains GCs exhibiting strong Balmer
lines and A-type spectra, from which one infers that these GCs must be
very young. \citet{Beasley04} and \citet{puzia05} confirmed this
conclusion of \citet{bh00}. \citet{Burstein04} and \citet{Fusi05}
increased the sample of young M31 GCs to 67. Very recently,
\citet{caldwell08} determined the ages and reddening values of 140
young clusters in M31 by comparing the observed spectra with model
spectra, and these clusters are less than 2 Gyr old. Most have ages
between $10^8$ and $10^9$ yr. The ages of the M31 clusters determined
in this paper are in agreement with previous determinations.

Many M31 GCs are resolved in {\sl HST} observations. Some authors,
including \citet{Ajhar96}, \citet{fusi96}, \citet{rich96},
\citet{hfr97}, \citet{Jablonka00}, \citet{Williams01a},
\citet{Williams01b}, and \citet{rich05}, used WFPC2 images to
construct CMDs for determination of the clusters' metallicities,
reddening values, and ages. However, these CMDs are usually not deep
enough to show conspicuous MSTOs and thus be useful for robust age
determinations. In fact, only \citet{Williams01a} and
\citet{Williams01b} managed to estimate the ages for four blue
massive, compact star clusters and 79 candidate young star clusters
by fitting isochrones to the stellar photometry.

Our sample contains four GCs in common with \citet{Ajhar96} (B006 and
B045), \citet{fusi96} (also B006 and B045), and \citet{rich05} (B012
and B233). However, only B012 \citep{rich05} could be older than about
8 Gyr (see Gallart et al. 2005), given the presence of a prominent
blue horizontal branch, which compares rather poorly with the age
obtained here ($\sim2$ Gyr). However, even with multi-passband
photometry spanning from the FUV to $K_s$, we can only determine
cluster ages in a {\it statistical} sense \citep[also
see][]{GG02,degrijs05}.

This discrepancy of a few Gyr highlights the difficulty of obtaining
age estimates of unresolved intermediate-age clusters based on
multi-passband photometry, given that the color evolution of SSPs is
only minimally age dependent once the population has reached an age of
$\sim 1 - 3$ Gyr. In addition, although the general age distribution
of an entire cluster population (in a given galaxy) can be retrieved
fairly self-consistently, the clusters' individual age determinations
depend rather strongly on the approach taken to fitting their ages
\citep[cf.][]{degrijs05}. This, combined with a strong dependence on
the adopted reddening and metallicity, results in {\it individual} age
estimates of intermediate-age clusters associated with large
uncertainties.

Cluster ages can also be derived by comparing observed with model
spectra.  Cross-identification of \citet{Beasley04}, \citet{puzia05},
and \citet{caldwell08} with the sample in this paper reveals that only
six GCs (and GC candidates) overlap with \citet{puzia05} (B006, B012,
B045, and B232) and \citet{caldwell08} (B049 and B195). The age
deerminations of two of these (B012 and B232) are inconsistent between
\citet{puzia05} and this paper: $10.2\pm2.9$ Gyr and $9.0\pm3.3$ Gyr
obtained by \citet{puzia05} compared to $2.0\pm0.1$ Gyr and
$2.0\pm0.1$ Gyr, respectively, obtained in this paper. The ages of the
other GCs are consistent among the different determinations.

In fact, the ages of GCs derived by different authors based on a
range of methods are not always consistent. For example, the ages of
B292 and B327 derived by \citet{Beasley04} and \citet{puzia05} are
$2.748\pm1.151$ Gyr and $0.080\pm0.929$ Gyr \citep{Beasley04}
compared to $9.2\pm3.3$ Gyr and $5.4\pm1.4$ Gyr \citep{puzia05},
respectively. On the other hand, \citet{caldwell08} estimated the
age of B327 at $0.050$ Gyr. In addition, the ages of clusters
derived in the same paper but based on different line-index
measurements are not always consistent either and may indeed differ
significantly \citep[see the upper panels of Fig. 5 in][]{puzia05}.

\section*{Acknowledgments}
We are indebted to the referee for thoughtful comments and
insightful suggestions that improved this paper greatly. This study
has been supported by the Chinese National Natural Science
Foundation through grants 10873016, 10803007, 10473012, 10573020,
10633020, 10673012, and 10603006, and by National Basic Research
Program of China (973 Program), No. 2007CB815403.

\newpage
\begin{table}
\begin{center}
\caption{BATC filter parameters and observational statistics}
\label{t1.tab}
\begin{tabular}{cccccc}
\tableline\tableline Filter & $\lambda_{\rm central}$ (\AA) & FWHM
(\AA) & N$^a$&
Exp.$^b$ & rms$^c$ \\
\tableline
 $c$ & 4210 & 320 &  3 & 01:00&0.015\\
 $d$ & 4540 & 340 &  3 & 01:00&0.009\\
 $e$ & 4925 & 390 &  3 & 01:00&0.015\\
 $f$ & 5270 & 340 &  3 & 01:00&0.006\\
 $g$ & 5795 & 310 &  3 & 01:00&0.003\\
 $h$ & 6075 & 310 &  3 & 01:00&0.003\\
 $i$ & 6656 & 480 &  3 & 01:00&0.003\\
 $j$ & 7057 & 300 &  3 & 01:00&0.008\\
 $k$ & 7546 & 330 &  3 & 01:00&0.004\\
 $m$ & 8023 & 260 &  3 & 01:00&0.003\\
 $n$ & 8480 & 180 &  6 & 02:00&0.004\\
 $o$ & 9182 & 260 &  6 & 02:00&0.003\\
 $p$ & 9739 & 270 &  6 & 02:00&0.009\\
\tableline
\end{tabular}\\
\vspace{3mm}
{$^a$ Number of exposures for each BATC filter\\}

{$^b$ Exposure time (in hr:min)}

{$^c$Zero-point error (in mag)}
\end{center}
\end{table}

\newpage
\begin{table}
\begin{center}
\caption{BATC intermediate-band photometry of our sample of 39 GCs and GC
candidates in M31.}
\label{t2.tab}
\begin{tabular}{cccccccccccccc}
\tableline\tableline
Name & $c$& $d$& $e$&$f$ &$g$ &$h$ &$i$ &$j$  &$k$ &$m$  &$n$  &$o$& $p$ \\
\hline
B004  & 17.71& 17.49& 17.22& 17.06& 16.72& 16.61& 16.44& 16.34& 16.21& 16.03& 16.09& 15.91& 16.01\\
      & 0.130& 0.012& 0.008& 0.009& 0.009& 0.008& 0.008& 0.009& 0.010& 0.009& 0.021& 0.010& 0.024\\
B006  & 16.68& 16.17& 15.84& 15.67& 15.29& 15.25& 15.09& 14.98& 14.82& 14.67& 14.70& 14.58& 14.53\\
      & 0.009& 0.005& 0.004& 0.004& 0.004& 0.004& 0.004& 0.003& 0.004& 0.003& 0.008& 0.005& 0.008\\
B008  & 17.83& 17.33& 17.04& 16.88& 16.51& 16.43& 16.26& 16.14& 16.01& 15.87& 15.73& 15.80& 15.73\\
      & 0.016& 0.010& 0.007& 0.008& 0.008& 0.007& 0.008& 0.008& 0.009& 0.007& 0.015& 0.011& 0.019\\
B010  & 17.48& 17.21& 16.94& 16.79& 16.50& 16.44& 16.25& 16.17& 16.10& 15.95& 16.00& 15.83& 15.82\\
      & 0.014& 0.010& 0.007& 0.008& 0.008& 0.007& 0.008& 0.008& 0.011& 0.008& 0.019& 0.012& 0.023\\
B012  & 15.90& 15.58& 15.34& 15.20& 14.91& 14.85& 14.71& 14.63& 14.55& 14.40& 14.45& 14.36& 14.35\\
      & 0.006& 0.004& 0.003& 0.003& 0.003& 0.003& 0.003& 0.003& 0.004& 0.003& 0.007& 0.004& 0.007\\
B013  & 18.35& 17.86& 17.51& 17.36& 16.99& 16.92& 16.75& 16.64& 16.47& 16.33& 16.36& 16.24& 16.19\\
      & 0.022& 0.012& 0.009& 0.010& 0.010& 0.010& 0.009& 0.010& 0.012& 0.009& 0.029& 0.015& 0.027\\
B016  & 18.64& 18.08& 17.78& 17.62& 17.24& 17.15& 16.95& 16.82& 16.68& 16.51& 16.56& 16.33& 16.33\\
      & 0.027& 0.016& 0.011& 0.012& 0.012& 0.011& 0.011& 0.010& 0.014& 0.010& 0.028& 0.016& 0.029\\
B019  & 15.75& 15.45& 15.15& 14.99& 14.61& 14.53& 14.38& 14.25& 14.11& 13.96& 13.89& 13.80& 13.76\\
      & 0.006& 0.004& 0.003& 0.003& 0.003& 0.003& 0.003& 0.003& 0.003& 0.003& 0.006& 0.004& 0.007\\
B020  & 15.61& 15.29& 15.06& 14.92& 14.54& 14.48& 14.36& 14.24& 14.12& 13.95& 13.99& 13.92& 13.85\\
      & 0.006& 0.003& 0.003& 0.003& 0.003& 0.003& 0.003& 0.003& 0.003& 0.003& 0.006& 0.004& 0.007\\
B022  & 17.96& 17.74& 17.53& 17.40& 17.09& 17.05& 16.96& 16.89& 16.80& 16.65& 16.65& 16.59& 16.61\\
      & 0.022& 0.016& 0.013& 0.014& 0.014& 0.013& 0.015& 0.016& 0.022& 0.017& 0.042& 0.028& 0.047\\
B026  & 18.58& 18.09& 17.70& 17.55& 17.15& 17.06& 16.88& 16.72& 16.54& 16.37& 16.27& 16.18& 16.19\\
      & 0.031& 0.023& 0.016& 0.018& 0.020& 0.015& 0.017& 0.015& 0.018& 0.018& 0.033& 0.022& 0.035\\
B035  & 18.45& 17.98& 17.68& 17.55& 17.17& 17.06& 16.92& 16.81& 16.62& 16.49& 16.52& 16.39& 16.33\\
      & 0.026& 0.015& 0.011& 0.011& 0.012& 0.011& 0.011& 0.011& 0.013& 0.011& 0.030& 0.018& 0.028\\
B045  & 16.88& 16.41& 16.08& 15.94& 15.55& 15.48& 15.30& 15.19& 15.06& 14.92& 14.93& 14.77& 14.75\\
      & 0.010& 0.006& 0.004& 0.005& 0.005& 0.004& 0.004& 0.004& 0.005& 0.004& 0.009& 0.007& 0.010\\
B047  & 18.36& 17.95& 17.68& 17.59& 17.23& 17.15& 17.02& 16.96& 16.86& 16.69& 16.77& 16.70& 16.73\\
      & 0.025& 0.015& 0.011& 0.011& 0.013& 0.011& 0.012& 0.012& 0.016& 0.015& 0.038& 0.023& 0.035\\
B049  & 17.93& 17.76& 17.70& 17.61& 17.43& 17.42& 17.30& 17.25& 17.16& 17.12& 17.28& 16.92& 17.92\\
      & 0.025& 0.022& 0.020& 0.021& 0.024& 0.019& 0.021& 0.024& 0.026& 0.021& 0.061& 0.040& 0.412\\
B050  & 17.68& 17.32& 17.03& 16.87& 16.49& 16.42& 16.27& 16.14& 16.01& 15.84& 15.78& 15.70& 15.73\\
      & 0.020& 0.014& 0.011& 0.010& 0.012& 0.010& 0.011& 0.011& 0.013& 0.012& 0.026& 0.017& 0.028\\
B052  & 19.02& 18.24& 17.64& 17.27& 16.92& 16.76& 16.53& 16.38& 16.23& 16.04& 15.89& 15.73& 15.65\\
      & 0.046& 0.020& 0.012& 0.011& 0.012& 0.010& 0.009& 0.009& 0.012& 0.009& 0.020& 0.013& 0.021\\
B062  & 18.86& 18.17& 17.66& 17.24& 16.96& 16.74& 16.52& 16.44& 16.31& 16.13& 15.98& 15.83& 15.79\\
      & 0.042& 0.021& 0.014& 0.010& 0.013& 0.010& 0.011& 0.011& 0.014& 0.011& 0.026& 0.016& 0.024\\
B074  & 17.55& 17.16& 16.91& 16.80& 16.43& 16.35& 16.22& 16.14& 16.03& 15.89& 15.92& 15.84& 15.80\\
      & 0.014& 0.010& 0.007& 0.007& 0.008& 0.007& 0.008& 0.008& 0.009& 0.008& 0.027& 0.013& 0.020\\
B081  & 17.05& 17.01& 16.87& 16.81& 16.59& 16.42& 16.44& 16.18& 16.08& 16.01& 15.95& 15.53& 15.86\\
      & 0.016& 0.015& 0.013& 0.013& 0.015& 0.014& 0.024& 0.034& 0.032& 0.029& 0.040& 0.049& 0.107\\
B089  & 18.24& 18.22& 18.16& 18.16& 18.11& 18.16& 18.16& 18.16& 18.07& 18.01& 18.49& 18.17& 18.34\\
      & 0.027& 0.026& 0.026& 0.024& 0.039& 0.034& 0.041& 0.045& 0.050& 0.047& 0.126& 0.090& 0.163\\
\tableline
\end{tabular}
\end{center}
\end{table}
\addtocounter{table}{-1}

\begin{table}
\begin{center}
\caption{Continued.} \label{t2.tab}
\begin{tabular}{cccccccccccccc}
\tableline\tableline
Name & $c$& $d$& $e$&$f$ &$g$ &$h$ &$i$ &$j$  &$k$ &$m$  &$n$  &$o$& $p$ \\
\hline
B100  & 18.82& 18.31& 18.04& 17.92& 17.63& 17.61& 17.31& 17.25& 17.24& 17.21& 17.24& 16.95& 16.91\\
      & 0.050& 0.030& 0.026& 0.026& 0.027& 0.025& 0.026& 0.048& 0.031& 0.033& 0.065& 0.037& 0.170\\
B129  & ...  & ...  & 18.41& 17.71& 16.83& 16.59& 16.07& 15.79& 15.47& 15.21& 14.99& 14.71& 14.62\\
      & ...  & ...  & 0.090& 0.059& 0.034& 0.030& 0.022& 0.019& 0.016& 0.014& 0.014& 0.012& 0.014\\
B156  & 17.67& 17.32& 17.09& 16.82& 16.53& 16.64& 16.44& 16.35& 16.17& 16.20& 16.18& 16.13& 16.40\\
      & 0.018& 0.011& 0.008& 0.008& 0.008& 0.010& 0.012& 0.014& 0.012& 0.031& 0.021& 0.035& 0.008\\
B168  & 19.50& 18.76& 18.23& 17.96& 17.29& 17.13& 16.78& 16.60& 16.28& 16.07& 15.95& 15.80& 15.69\\
      & 0.083& 0.053& 0.030& 0.026& 0.025& 0.017& 0.015& 0.015& 0.014& 0.010& 0.027& 0.013& 0.020\\
B170  & 18.27& 17.90& 17.61& 17.46& 17.11& 17.06& 16.84& 16.76& 16.66& 16.52& 16.46& 16.38& 16.44\\
      & 0.025& 0.020& 0.012& 0.013& 0.014& 0.012& 0.012& 0.014& 0.015& 0.014& 0.033& 0.018& 0.141\\
B195  & 18.97& 18.78& 18.61& 18.57& 18.38& 18.36& 18.35& 18.11& 18.20& 18.04& 17.96& 17.96& ...\\
      & 0.046& 0.035& 0.028& 0.035& 0.044& 0.033& 0.044& 0.044& 0.062& 0.059& 0.109& 0.074& ...\\
B199  & 18.27& 18.00& 17.80& 17.62& 17.44& 17.41& 17.22& 17.10& 17.06& 17.00& 17.05& 16.87& 17.06\\
      & 0.024& 0.018& 0.012& 0.013& 0.016& 0.013& 0.014& 0.016& 0.018& 0.017& 0.039& 0.022& 0.060\\
B207  & 18.04& 17.74& 17.53& 17.36& 17.14& 17.10& 16.93& 16.84& 16.81& 16.73& 16.71& 16.59& 16.64\\
      & 0.020& 0.014& 0.011& 0.012& 0.014& 0.013& 0.014& 0.015& 0.020& 0.015& 0.039& 0.023& 0.053\\
B212  & 16.17& 15.91& 15.69& 15.51& 15.31& 15.28& 15.10& 15.01& 14.96& 14.91& 14.84& 14.76& 14.80\\
      & 0.007& 0.005& 0.004& 0.004& 0.004& 0.004& 0.004& 0.004& 0.005& 0.004& 0.011& 0.006& 0.012\\
B219  & 17.25& 16.90& 16.63& 16.44& 16.15& 16.05& 15.88& 15.76& 15.62& 15.46& 15.46& 15.32& 15.33\\
      & 0.013& 0.008& 0.006& 0.006& 0.007& 0.006& 0.006& 0.006& 0.007& 0.006& 0.014& 0.008& 0.018\\
B226  & 19.13& 18.51& 18.10& 17.77& 17.55& 17.43& 17.15& 17.05& 16.97& 16.82& 16.63& 16.56& 16.49\\
      & 0.044& 0.029& 0.015& 0.015& 0.018& 0.015& 0.016& 0.016& 0.022& 0.016& 0.041& 0.021& 0.045\\
B230  & 16.44& 16.33& 16.13& 15.98& 15.75& 15.68& 15.56& 15.47& 15.41& 15.21& 15.21& 15.20& 15.20\\
      & 0.009& 0.006& 0.005& 0.005& 0.006& 0.005& 0.005& 0.006& 0.007& 0.005& 0.015& 0.008& 0.020\\
B232  & 16.15& 16.02& 15.80& 15.65& 15.34& 15.28& 15.17& 15.06& 14.98& 14.80& 14.74& 14.76& 14.71\\
      & 0.008& 0.005& 0.004& 0.005& 0.005& 0.005& 0.005& 0.005& 0.006& 0.006& 0.011& 0.008& 0.016\\
B233  & 16.41& 16.19& 15.94& 15.81& 15.44& 15.38& 15.26& 15.15& 15.01& 14.85& 14.85& 14.73& 14.90\\
      & 0.010& 0.008& 0.007& 0.006& 0.007& 0.006& 0.007& 0.006& 0.008& 0.007& 0.010& 0.015& 0.019\\
B236  & 17.86& 17.69& 17.52& 17.38& 17.08& 17.03& 16.88& 16.75& 16.47& 16.40& 16.45& 16.76& 16.22\\
      & 0.021& 0.017& 0.012& 0.012& 0.015& 0.012& 0.012& 0.018& 0.023& 0.040& 0.021& 0.039& 0.164\\
B237  & 17.92& 17.70& 17.45& 17.31& 17.01& 16.94& 16.79& 16.70& 16.63& 16.48& 16.50& 16.42& 16.47\\
      & 0.020& 0.015& 0.012& 0.012& 0.015& 0.013& 0.015& 0.015& 0.016& 0.018& 0.033& 0.024& 0.145\\
B238  & 17.39& 17.02& 16.73& 16.58& 16.23& 16.13& 15.97& 15.88& 15.74& 15.66& 15.58& 15.58& 15.45\\
      & 0.014& 0.007& 0.006& 0.006& 0.007& 0.012& 0.007& 0.016& 0.016& 0.014& 0.019& 0.011& 0.055\\
B239  & 17.88& 17.79& 17.51& 17.43& 17.01& 16.90& 16.77& 16.66& 16.52& 16.43& 16.33& 16.32& 16.15\\
      & 0.156& 0.062& 0.040& 0.037& 0.033& 0.025& 0.017& 0.032& 0.032& 0.026& 0.037& 0.063& 0.103\\
\tableline
\end{tabular}
\end{center}
\end{table}

\begin{table}
\begin{center}
\caption{GALEX, broad-band, and 2MASS photometry of the 39 M31 GCs
and GC candidates.}
\label{t3.tab}
\begin{tabular}{cccccccccccc}
\tableline\tableline
Name & $c$\tablenotemark{\dag}  &   FUV & NUV & $U$ & $B$ & $V$ & $R$ & $I$ & $J$ & $H$  & $K_s$ \\
\hline
B004 & 1    & ... & 22.25& 18.29& 17.87& 16.95& 16.36& 15.73& 14.91& 14.36& 14.19\\
      && ... &  0.07&  0.03&  0.01&  0.01&  0.02&  0.01&  0.02&  0.07&  0.05\\
B006 & 1    & ... & 21.41& 16.94& 16.49& 15.53& 14.97& 14.31& 13.48& 12.75& 12.61\\
      && ... &  0.04&  0.02&  0.01&  0.01&  0.01&  0.01&  0.03&  0.03&  0.03\\
B008 & 1    & ... & 22.59& 18.16& 17.66& 16.56& 16.21& 15.51& 14.68& 14.17& 13.98\\
      && ... &  0.12&  0.08&  0.05&  0.05&  0.05&  0.05&  0.05&  0.07&  0.06\\
B010 & 1    & 21.93& 20.87& 17.65& 17.50& 16.66& 16.12& 15.48& 14.76& 14.41& 14.07\\
      &&  0.08&  0.03&  0.02&  0.01&  0.01&  0.01&  0.01&  0.03&  0.07&  0.06\\
B012 & 1    & 20.10& 19.02& 15.99& 15.86& 15.13& 14.62& 14.08& 13.36& 12.79& 12.72\\
      &&  0.02&  0.01&  0.01&  0.01&  0.01&  0.01&  0.01&  0.03&  0.03&  0.03\\
B013 & 1    & ... & ... & 18.56& 18.06& 17.19& 16.60& 15.96& 15.18& 14.61& 14.34\\
      && ... & ... &  0.05&  0.02&  0.01&  0.02&  0.02&  0.03&  0.07&  0.05\\
B016 & 1    & ... & ... & 18.86& 18.58& 17.58& 16.85& 16.15& 15.18& 14.18& 14.05\\
      && ... & ... &  0.08&  0.04&  0.01&  0.03&  0.02&  0.08&  0.07&  0.10\\
B019 & 1    & 20.81& 19.97& 16.36& 15.94& 14.93& 14.31& 13.74& 12.86& 12.10& 11.96\\
      &&  0.04&  0.02&  0.01&  0.01&  0.01&  0.05&  0.01&  0.02&  0.03&  0.02\\
B020 & 1    & 20.05& 19.20& 15.98& 15.74& 14.91& 14.37& 13.65& 12.97& 12.26& 12.21\\
      &&  0.02&  0.01&  0.08&  0.05&  0.05&  0.05&  0.05&  0.02&  0.03&  0.03\\
B022 & 1    & 22.69& 21.20& 18.14& 18.09& 17.36& 16.97& 16.35& 15.75& 15.04& 15.48\\
      &&  0.16&  0.04&  0.08&  0.02&  0.01&  0.02&  0.02&  0.08&  0.10&  0.12\\
B026 & 1    & ... & ... & 19.14& 18.60& 17.53& 16.88& 16.22& 15.10& 14.48& 14.00\\
      && ... & ... &  0.06&  0.02&  0.01&  0.03&  0.02&  0.08&  0.07&  0.05\\
B035 & 1    & ... & 22.61& 18.52& 18.37& 17.48& 16.81& 16.24& 15.30& 14.67& 14.42\\
      && ... &  0.10&  0.08&  0.03&  0.01&  0.02&  0.02&  0.08&  0.07&  0.10\\
B045 & 1    & ... & 21.07& 17.09& 16.72& 15.78& 15.19& 14.54& 13.73& 13.00& 12.89\\
      && ... &  0.03&  0.02&  0.01&  0.01&  0.01&  0.01&  0.03&  0.04&  0.03\\
B047 & 1    & 22.68& 21.30& 18.32& 18.23& 17.51& 16.88& 16.30& 15.86& 15.24& 15.47\\
      &&  0.15&  0.04&  0.06&  0.02&  0.01&  0.03&  0.02&  0.08&  0.10&  0.12\\
B049 & 1    & ... & 21.55& 18.26& 18.08& 17.56& 17.11& 16.87& 15.61& 15.33& 14.68\\
      && ... &  0.06&  0.09&  0.04&  0.01&  0.04&  0.04&  0.08&  0.10&  0.10\\
B050 & 1    & ... & 22.18& 18.09& 17.76& 16.84& 16.27& 15.66& 14.72& 14.22& 13.96\\
      && ... &  0.09&  0.05&  0.02&  0.01&  0.02&  0.01&  0.03&  0.07&  0.05\\
B052 & 4    & ... & ... & 19.80& 18.62& 17.21& 16.54& 15.77& 14.70& 14.01& 13.40\\
      && ... & ... &  0.08&  0.02&  0.01&  0.02&  0.02&  0.04&  0.07&  0.05\\
B062 & 4    & ... & ... & 19.33& 18.58& 17.24& 16.61& 15.82& 14.89& 14.21& 13.66\\
      && ... & ... &  0.08&  0.02&  0.01&  0.02&  0.01&  0.04&  0.07&  0.05\\
B074 & 1    & 22.12& 20.75& 17.54& 17.40& 16.65& 16.14& 15.58& 14.83& 13.95& 14.11\\
      &&  0.07&  0.02&  0.03&  0.01&  0.01&  0.01&  0.01&  0.02&  0.04&  0.04\\
B081 & 1    & 21.73& 20.47& 17.60& 17.34& 16.80& 16.36& 15.73& 14.82& 14.01& 13.96\\
      &&  0.13&  0.04&  0.02&  0.01&  0.01&  0.02&  0.02&  0.03&  0.07&  0.05\\
B089 & 2    & 19.89& 19.62& 17.96& 18.28& 18.18& 18.22& 17.70& ... & ... & ... \\
      &&  0.03&  0.02&  0.05&  0.04&  0.03&  0.06&  0.06& ... & ... & ... \\
\tableline
\end{tabular}
\end{center}
\end{table}
\addtocounter{table}{-1}

\begin{table}
\begin{center}
\caption{Continued.} \label{t3.tab}
\begin{tabular}{cccccccccccc}
\tableline\tableline
Name & $c$  &  FUV & NUV & $U$ & $B$ & $V$ & $R$ & $I$ & $J$ & $H$  & $K_s$ \\
\hline
B100 & 1    & 22.37& ... & 18.94& 19.05& 17.91& ... & 17.77& 15.85& 14.70& 14.67\\
      &&  0.19& ... &  0.08&  0.07&  0.05& ... &  0.07&  0.08&  0.07&  0.10\\
B129 & 1    & ... & ... & ... & 19.56& 17.40& ... & 14.69& 13.25& 12.40& 12.19\\
      && ... & ... & ... &  0.05&  0.05& ... &  0.05&  0.03&  0.03&  0.04\\
B156 & 1    & ... & 20.99& 17.89& 17.63& 16.84& 16.37& 15.87& ... & ... & ... \\
      && ... &  0.09&  0.02&  0.02&  0.01&  0.02&  0.05& ... & ... & ... \\
B168 & 1    & ... & ... & 20.82& 19.23& 17.63& 16.69& 15.72& 14.52& 13.43& 13.37\\
      && ... & ... &  0.08&  0.06&  0.01&  0.03&  0.02&  0.06&  0.04&  0.08\\
B170 & 1    & ... & ... & 18.90& 18.37& 17.39& 16.80& 16.17& 15.38& 14.75& 14.61\\
      && ... & ... &  0.06&  0.02&  0.01&  0.02&  0.02&  0.08&  0.07&  0.10\\
B195 & 2    & ... & ... & 19.94& 18.97& 18.57& 18.02& 17.59& ... & ... & ... \\
      && ... & ... &  0.08&  0.06&  0.01&  0.07&  0.05& ... & ... & ... \\
B199 & 1    & ... & 21.53& 18.45& 18.37& 17.60& 17.03& 16.57& 16.06& 15.42& 15.39\\
      && ... &  0.10&  0.08&  0.03&  0.01&  0.03&  0.02&  0.10&  0.10&  0.12\\
B207 & 1    & 21.64& 21.04& 18.26& 18.07& 17.33& 16.81& 16.33& 15.67& 14.42& 14.78\\
      &&  0.11&  0.06&  0.03&  0.02&  0.01&  0.02&  0.02&  0.05&  0.07&  0.08\\
B212 & 1    & 20.27& 19.16& 16.23& 16.22& 15.48& 15.00& 14.48& 13.82& 13.17& 13.11\\
      &&  0.05&  0.02&  0.02&  0.01&  0.01&  0.01&  0.01&  0.03&  0.04&  0.05\\
B219 & 1    & ... & 21.59& 17.74& 17.32& 16.39& 15.82& 15.19& 14.32& 13.71& 13.51\\
      && ... &  0.10&  0.08&  0.05&  0.05&  0.05&  0.05&  0.02&  0.04&  0.04\\
B226 & 2    & 22.09& 21.61& 19.08& 19.04& 17.65& ... & 16.32& 15.21& 14.47& 14.14\\
      &&  0.14&  0.09&  0.08&  0.05&  0.05& ... &  0.05&  0.08&  0.07&  0.10\\
B230 & 1    & 20.69& 19.46& 16.78& 16.77& 16.05& 15.61& 15.13& 14.43& 13.92& 13.85\\
      &&  0.08&  0.03&  0.02&  0.01&  0.01&  0.01&  0.01&  0.02&  0.04&  0.05\\
B232 & 1    & 20.67& 19.49& 16.53& 16.38& 15.70& 15.20& 14.65& 13.94& 13.36& 13.25\\
      &&  0.06&  0.02&  0.01&  0.01&  0.01&  0.01&  0.01&  0.03&  0.04&  0.05\\
B233 & 1    & 21.27& 20.04& 16.82& 16.61& 15.80& 15.27& 14.76& 13.90& 13.32& 13.21\\
      &&  0.05&  0.01&  0.01&  0.01&  0.01&  0.01&  0.01&  0.02&  0.04&  0.03\\
B236 & 1    & ... & 21.07& 18.21& 18.20& 17.38& 16.97& 16.24& ... & ... & ... \\
      && ... &  0.07&  0.08&  0.02&  0.01&  0.02&  0.02& ... & ... & ... \\
B237 & 1    & ... & 21.25& 18.03& 17.87& 17.10& 16.57& 16.05& 15.47& 15.06& 14.91\\
      && ... &  0.05&  0.02&  0.02&  0.01&  0.02&  0.02&  0.04&  0.10&  0.09\\
B238 & 1    & ... & 21.91& 17.73& 17.39& 16.42& 15.86& 15.22& 14.46& 13.72& 13.67\\
      && ... &  0.08&  0.02&  0.01&  0.01&  0.01&  0.01&  0.04&  0.04&  0.05\\
B239 & 1    & ... & ... & 18.49& 18.10& 17.08& 16.65& 16.09& 15.31& 14.55& 14.55\\
      && ... & ... &  0.04&  0.02&  0.01&  0.02&  0.02&  0.04&  0.07&  0.08\\
\tableline
\end{tabular}\\
\vspace{3mm}
{$\dag$ New classification flag, following RBC V3.5
notation. 1 = confirmed GC, 2 = GC candidate, 4 = confirmed galaxy}
\end{center}
\end{table}

\begin{table}
\begin{center}
\caption{Reddening values and metallicities for our 39 M31 GCs and
GC candidates.} \label{t4.tab}
\begin{tabular}{ccccc}
\tableline \tableline
Name & $E(B-V)$& ref.$^a$ & \multicolumn{1}{c}{$\rm [Fe/H]$}& ref.$^b$\\
\hline
B004  &   0.07$\pm$   0.02&  1&  $-0.31\pm$   0.74&  1\\
B006  &   0.09$\pm$   0.02&  1&  $-0.58\pm$   0.10&  1\\
B008  &   0.21&  2&              $-0.41\pm$   0.38&  1\\
B010  &   0.22$\pm$   0.01&  1&  $-1.77\pm$   0.14&  1\\
B012  &   0.12$\pm$   0.01&  1&  $-1.65\pm$   0.19&  1\\
B013  &   0.13$\pm$   0.02&  1&  $-1.01\pm$   0.49&  1\\
B016  &   0.30$\pm$   0.02&  1&  $-0.78\pm$   0.19&  1\\
B019  &   0.20$\pm$   0.01&  1&  $-1.09\pm$   0.02&  1\\
B020  &   0.12$\pm$   0.01&  1&  $-1.07\pm$   0.10&  3\\
B022  &   0.04$\pm$   0.03&  1&  $-1.64\pm$   0.07&  4\\
B026  &   0.15$\pm$   0.02&  1&   $0.01\pm$   0.38&  1\\
B035  &   0.27$\pm$   0.05&  2&  $-0.20\pm$   0.54&  1\\
B045  &   0.18$\pm$   0.01&  1&  $-1.05\pm$   0.25&  1\\
B047  &   0.09$\pm$   0.02&  1&  $-1.62\pm$   0.41&  1\\
B049  &   0.16$\pm$   0.02&  1&  $-2.14\pm$   0.55&  1\\
B050  &   0.24$\pm$   0.01&  1&  $-1.42\pm$   0.37&  1\\
B052  &   0.23$\pm$   0.04&  1&   $0.12\pm$   0.17&  4\\
B062  &   0.26$\pm$   0.03&  1&  $-0.47\pm$   0.11&  4\\
B074  &   0.19$\pm$   0.01&  1&  $-1.88\pm$   0.06&  1\\
B081  &   0.11$\pm$   0.02&  1&  $-1.74\pm$   0.40&  1\\
B089  &   ...    &  ...&  ... &  ...                  \\
B100  &   0.48$\pm$   0.08&  1&  $-2.21\pm$   0.10&  4\\
B129  &   1.16$\pm$   0.06&  1&  $-1.21\pm$   0.32&  1\\
B156  &   0.10$\pm$   0.02&  1&  $-1.51\pm$   0.38&  1\\
B168  &   0.54$\pm$   0.05&  1&  $-0.12\pm$   0.21&  4\\
B170  &   0.10$\pm$   0.02&  1&  $-0.54\pm$   0.24&  1\\
B195  &   0.12$\pm$   0.00&  1&  $-1.48\pm$   0.63&  4\\
B199  &   0.10$\pm$   0.02&  1&  $-1.59\pm$   0.11&  1\\
B207  &   0.05$\pm$   0.02&  2&  $-0.81\pm$   0.59&  1\\
B212  &   0.13$\pm$   0.01&  1&  $-1.75\pm$   0.13&  3\\
B219  &   0.05$\pm$   0.03&  1&  $-0.01\pm$   0.57&  1\\
B226  &   1.08$\pm$   0.06&  1&  ...&...\\
B230  &   0.15$\pm$   0.01&  1&  $-2.17\pm$   0.16&  1\\
B232  &   0.14$\pm$   0.01&  1&  $-1.83\pm$   0.14&  1\\
B233  &   0.17$\pm$   0.01&  1&  $-1.59\pm$   0.32&  3\\
B236  &   0.07$\pm$   0.05&  1&  $-1.01\pm$   0.17&  4\\
B237  &   0.14$\pm$   0.02&  1&  $-2.09\pm$   0.28&  1\\
B238  &   0.11$\pm$   0.02&  1&  $-0.57\pm$   0.66&  1\\
B239  &   0.09$\pm$   0.01&  1&  $-1.18\pm$   0.61&  2\\
\tableline
\end{tabular}\\
\vspace{3mm}
{$^a$The reddening values are taken from \citet{fan08}
(ref.=1) and \citet{bh00} (ref.=2).\\} {$^b$The metallicities are
taken from \citet{per02} (ref.=1), \citet{bh00} (ref.=2),
\citet{hbk91} (ref.=3), and \citet{fan08} (ref.=4).}
\end{center}
\end{table}

\begin{table}
\begin{center}
\caption{Ages estimates for 35 GCs and GC candidates in M31.}
\label{t5.tab}
\begin{tabular}{ccc|ccc}
\tableline \tableline
Name &   Age &     $\chi_{\rm min}^2$ & Name &   Age &     $\chi_{\rm min}^2$\\
  & (Gyr) & (per degree of freedom)&  & (Gyr) & (per degree of freedom) \\
\hline
B004  & $  4.10\pm  0.55$ &  3.13&      B100  & $  0.50\pm  0.10$ & 14.38\\
B006  & $ 12.50\pm  0.65$ &  1.39&      B129  & $ 15.10\pm  0.70$ &  9.00\\
B008  & $  2.00\pm  0.10$ &  6.54&      B156  & $  4.90\pm  0.65$ &  2.09\\
B010  & $  1.80\pm  0.10$ &  1.08&      B168  & $ 12.60\pm  0.20$ &  3.27\\
B012  & $  2.00\pm  0.10$ &  1.54&      B170  & $  4.00\pm  0.45$ &  1.39\\
B013  & $ 12.00\pm  2.00$ &  0.96&      B195  & $  0.70\pm  0.15$ &  0.82\\
B016  & $  2.40\pm  0.30$ &  1.95&      B199  & $  3.30\pm  0.55$ &  1.38\\
B019  & $  2.10\pm  0.10$ & 12.01&      B207  & $  1.20\pm  0.10$ &  7.99\\
B020  & $  1.80\pm  0.10$ &  7.50&      B212  & $  1.80\pm  0.10$ &  0.75\\
B022  & $  3.40\pm  0.15$ &  4.11&      B219  & $  2.50\pm  0.15$ &  3.06\\
B026  & $  3.50\pm  0.25$ &  3.28&      B230  & $  1.60\pm  0.10$ &  4.59\\
B035  & $  1.00\pm  0.10$ &  3.97&      B232  & $  2.00\pm  0.10$ &  2.62\\
B045  & $  8.80\pm  1.45$ &  0.78&      B233  & $  2.30\pm  0.10$ &  3.73\\
B047  & $  2.80\pm  0.20$ &  4.22&      B236  & $  2.00\pm  0.25$ &  2.63\\
B049  & $  1.60\pm  0.10$ &  7.82&      B237  & $  3.50\pm  0.35$ &  1.41\\
B050  & $ 16.00\pm  0.30$ &  2.12&      B238  & $  5.00\pm  0.45$ &  2.01\\
B074  & $  2.10\pm  0.15$ &  2.12&      B239  & $ 14.50\pm  2.05$ &  1.70\\
B081  & $  2.10\pm  0.20$ &  7.82&            &                   &      \\
\tableline
\end{tabular}
\end{center}
\end{table}

\end{document}